# Galaxy Makers: *creating an online component to a science exhibition for re-engagement, evaluation and content legacy*

**Josh Borrow**

*Durham University MPhys Student, U.K., josh.borrow@gmail.com*

**Chris Harrison**

*European Southern Observatory Research Fellow, previously based at Department of Physics, Durham University, c.m.harrison@mail.com*

**Keywords:**

**Comsology, website, app, metrics, analytics, royal, society, re-engagement.**

**Abstract:**

For the Royal Society Summer Science Exhibition 2016, the Institute of Computational Cosmology from Durham University created the Galaxy Makers exhibit to communicate our computational cosmology and astronomy research. In addition to the physical exhibit we created an online component to foster re-engagement, create a permanent home for our content and to allow us to collect the ever-more important information about participation and impact. Here we summarise the details of the exhibit and the successes of creating an online component. We also share suggestions of further uses and improvements that could be implemented for online components of other science exhibitions.

## Introduction

Every summer, a carnival of science descends on London's Royal Society. Exhibitors from Universities and science organisations from around the UK are selected to bring their world-leading research to show off to the public at the Summer Science Exhibition. The one week long festival featured 22 curated exhibits as well as an extensive programme of talks and activities for all ages.

Our exhibit, *Galaxy Makers*, created by 50 members of staff and students from Durham University, showcased the wonderful imagery and science of a recent revolutionary set of computer simulated universes - the EAGLE simulations (Schaye et al 2015, see Figure 1). By using the dazzling images and movies produced in the simulation as essentially an advertising device, we can draw in the younger generation and capitalise on their love of all things digital. We can demonstrate that the same tools that are used to make their favourite video games also have a use in world-leading research and hopefully inspire them to pursue a computing-focussed career in scientific research or elsewhere. In an effort to inspire both young and old audiences we built Galaxy Makers as an interactive educational experience that combines science, technology, and art.

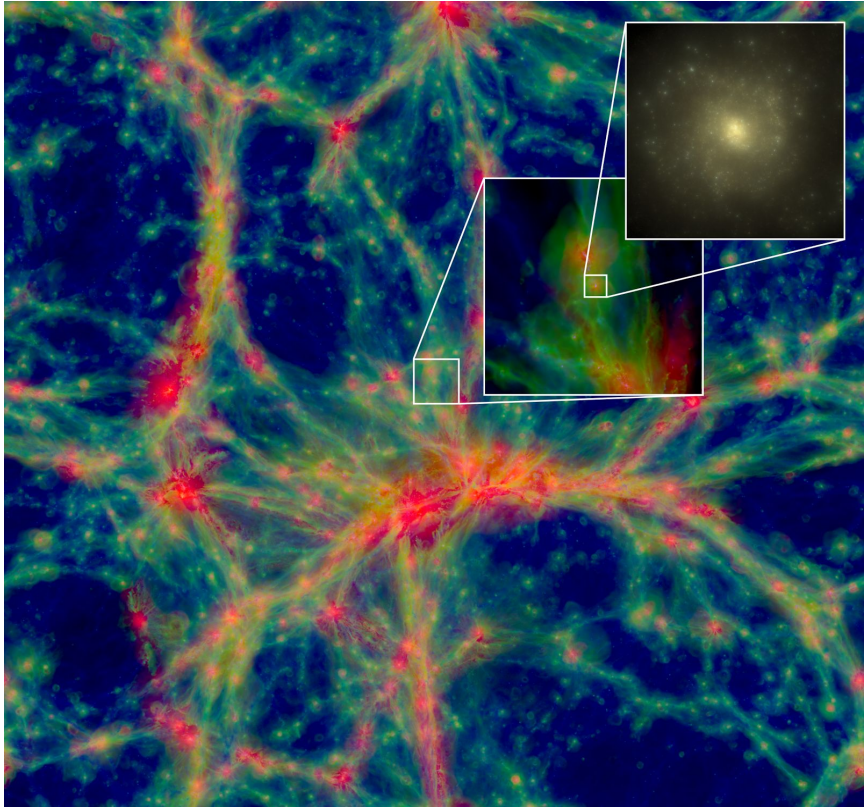

**Figure 1:** EAGLE (Evolution and Assembly of GaLaxies and their Environments) is a simulation aimed at understanding how galaxies form and evolve. The simulation took three months to run on a supercomputer and hence is highly detailed; the above picture shows a zoom-in from the 'cosmic web' to an individual galaxy.

## Why Communicate Computational Cosmology and Astronomy?

Computational Cosmology and Astronomy allow us to address fundamental questions about the Universe, from how it formed to its eventual fate. The question of origins is always relevant to the wider public. In other disciplines, scientists can run experiments to gain new information; however, we do not have this luxury - there is only one Universe! Therefore, computers became the laboratory bench of cosmologists and astronomers, in which simulated universes can be created and tested again and again. This allows cosmologists to develop and refine their ideas about the physics that shape galaxies. This is a great application of computational methods and a popular way to introduce them to the public. Furthermore, astronomical imagery is inspiring and captivating and is therefore incredibly useful for cultivating interest. Computer simulations can produce fantastic animations of the large-scale Universe that are impossible to recreate from observational data; we can watch how galaxies form, speeding up billions of years of cosmic evolution into just seconds.

Whilst the UK's school curriculum is finally changing to introduce programming, it is important that we keep "pushing from the top" to ensure that the public is informed that computational methods are a key part of research in many fields. It is not the case that all scientists wear a lab coat, wield a flask, and use a pencil all day, however, it appears that public opinion seems fairly centered within this view. Computational cosmology and astronomy lets us show the younger generation that in learning the initial concepts of programming in the classroom the eventual possibilities are endless.

# The Exhibition

The week-long in-person exhibition centered around the display and use of three machines (see below), which were available over 64 hours in total. The 14, 371 visitors to the exhibition included the media, school groups, families, invited guests (e.g. Fellows of the Royal Society), and other members of the public.

The machines were worked to nearly full capacity for the whole week, with the demonstrators (4-6 at a time) constantly engaged with participants. Through recording how many times each machine was used, we were able to establish that around 2000 people were able to use the machines, with more approaching the stand and engaging in conversation with the demonstrators about the project. Overall more than 3000 individuals were engaged with directly at the premier Galaxy Makers exhibition.

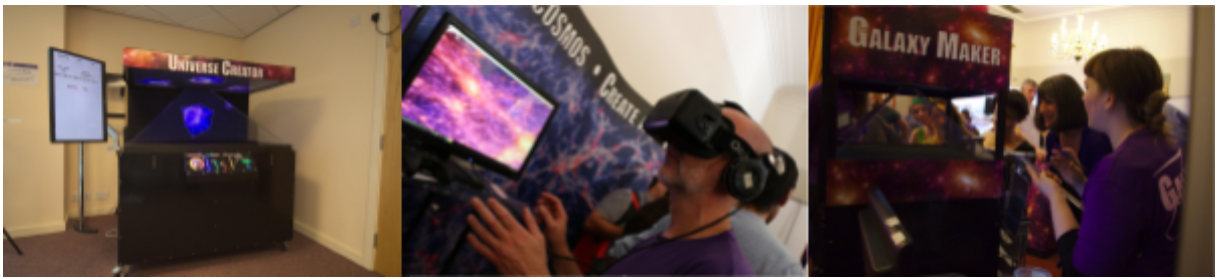

**Figure 2**: The three machines: the Universe Creator (left), Tour the Cosmos (center), and Galaxy Maker (right), that were presented at the Royal Society Summer Festival.

## The Galaxy Maker

What is a galaxy? Our Galaxy Maker activity (Figure 2) allows the participants to make their own "holographic" galaxy by following various galaxy model instructions (or "recipes") to weigh out the components (or "ingredients") of different galaxies. After measuring out these components and pouring them into the various chutes of the Galaxy Maker, a projected movie appears in a pyramid, giving the illusion of a hologram. Some of the messages we aim to portray with this activity are that galaxies are vast collections of stars, gas and dark matter and that galaxies come in different shapes and sizes; for example, our Milky Way is an example of a large spiral galaxy.

## Tour the Cosmos

What happens during the life of a galaxy? Our virtual reality Tour of the Cosmos (Figure 2) allows participants to submerge themselves right inside our computer simulation using an Oculus Rift. This unique experience flies the passengers through space and time whilst the audio commentary talks you through some of the key events of how galaxies are created.

## Create a Universe

How do we simulate galaxies? Our Universe Creator activity (Figure 2) allows participants to experience what it is like to run their own computer simulations. The participants use various knobs and levers to make their own assumptions about various aspects of astrophysics. For example, it is possible to change the amount of energy released by supermassive black holes and the amount of dark matter in the universe. Once these decisions have been made the computer simulation is projected inside a large pyramid, giving the illusion of a hologram. If these assumptions are unrealistic the participant may end up with a universe that is too hot or perhaps containing no galaxies at all! Participants will get to learn about how the experts have used similar experiments to understand how the Universe works.

## The Online Component

During the in-person exhibition, attendees who created a galaxy or universe on our machines were given small pyramids (Figure 3) that could be used to view the "hologram" movies at home on a smartphone or tablet through the Galaxy Makers website ([www.galaxymakers.org](www.galaxymakers.org), see Figure 4). Users were given a unique code that represented their universe/galaxy to type into the website. On entering the code, the user is provided with a movie to replay their "holographic" movie inside their pyramids. Additionally, further information is provided about their universe or galaxy so they can learn more about what they created. It is also possible to make new universes and galaxies on the website for those who wanted to continue learning or for members of the public who could not attend the exhibit in person.

The pyramids were an incredibly cost-effective way of ensuring that attendees re-engaged with our content. Google Analytics was set up on the website to analyse its usage; this showed that of the people that were given a pyramid, 40% (around 700 people) entered the website with 75% of those making repeat visits. In total, 898 unique users visited the website soon after or during the exhibition (as of July 19th 2016) from 22 different countries, with the majority of non-UK users being directed there through social media. Existing, related University websites also saw an increase in traffic during the exhibition period with the EAGLE project web pages having a 20% increase in pageviews.

Making the parallel online component of the exhibition required relatively little effort, requiring only two team members. However, the benefits were many: (1) we re-engaged with hundreds of people; (2) we ensured that our content has a meaningful, lasting presence; (3) we created educational resources for future use; (4) we were able to obtain vital statistics on participation and impact and (5) created online advertising for the institution. (6) we were able to engage people who could not visit Summer Science in person.

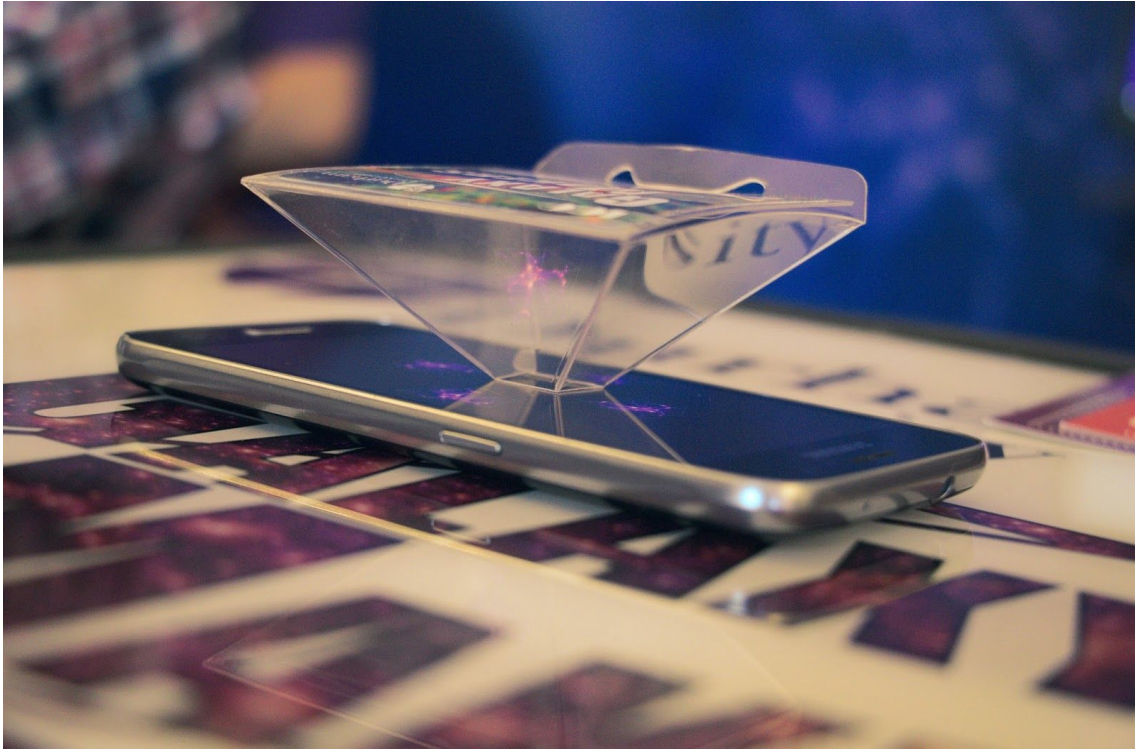

**Figure 3**: One of the 'hologram' pyramids in use with a smartphone. Note the 'floating' simulation - the effect is really something to behold in real life!

## Re-engagement

As well as reinforcing important concepts, re-engagement allows for deeper insights to be gained into the subject matter, particularly if the participant is interested. We aimed to achieve this with our website, by providing more information for interested readers through text and extra videos. This was particularly important for the Royal Society exhibition because participants only had a very limited time in which they could engage with the exhibition due to the extremely busy nature of the festival.

As many users visited the website more than once and hence engaged with the content at least 3 times, we are more than confident that we have made a lasting impression on hundreds of people that simply would not have been possible had we only utilised our in-person opportunity for engagement.

## A Record of Materials

As with all exhibitions, a large amount of effort was put into developing materials; in our case, this was the videos for the exhibition machines. It would have been a huge shame, both for the exhibitors and the public, for these materials to have only been available in one place at a specific time.

There are plans to show the Galaxy Makers exhibit at various points in the future but, even so, the range would have been extremely limited in comparison to a web-based exhibition. As our analytics have shown with our visitors originating from 22 countries, there is a worldwide audience for public engagement materials that should not be neglected.

## Further Education

Our online experience allows us to engage with a considerably larger audience in a much deeper way than the in-person exhibit alone, as we have previously discussed. In the future, we plan to produce educational resources for use by teachers with the Galaxy Makers website so that classroom users can engage with the content as well. Because the Galaxy Makers website was developed from the outset in parallel with the in-person exhibition, this is considerably more straightforward than it would have been if we needed to "convert" the exhibit into a classroom-friendly format.

## Measuring and Recording Impact

We live in an age where it is becoming more and more important to collect statistics and metrics to quantify the "impact" of our outreach activities. With an in-person exhibition, in particular when it is incredibly busy, it can be difficult to collect useful metrics and feedback. This means that public engagement evaluation reports can only consist of very simple, broad information such as numbers of members of the public reached. Online, this is a different story thanks to the off-the-shelf analytics programs like Google Analytics.

Google Analytics allows us to study user behaviour and flow, as well as more basic metrics such as the number of users on our site. Furthermore, this includes details of region/country of the visitor, their age brackets and number of times an individual visited the website. Whilst these metrics cannot simply be transplanted onto the in-person exhibition, they can give us key insights into the most popular parts, find out what users were most engaged by, and help track the long-term health of the exhibit.

As well as more information about the content of the exhibit, the Galaxy Makers website will allow us to catalogue the progress of the in-person exhibit with photos, updates and supplementary information about the exhibit that will be incredibly useful for when we are approaching other organisations or simply discussing our experience. The website allows us to give our metrics immediate context through the associated images and online exhibits that are available there.

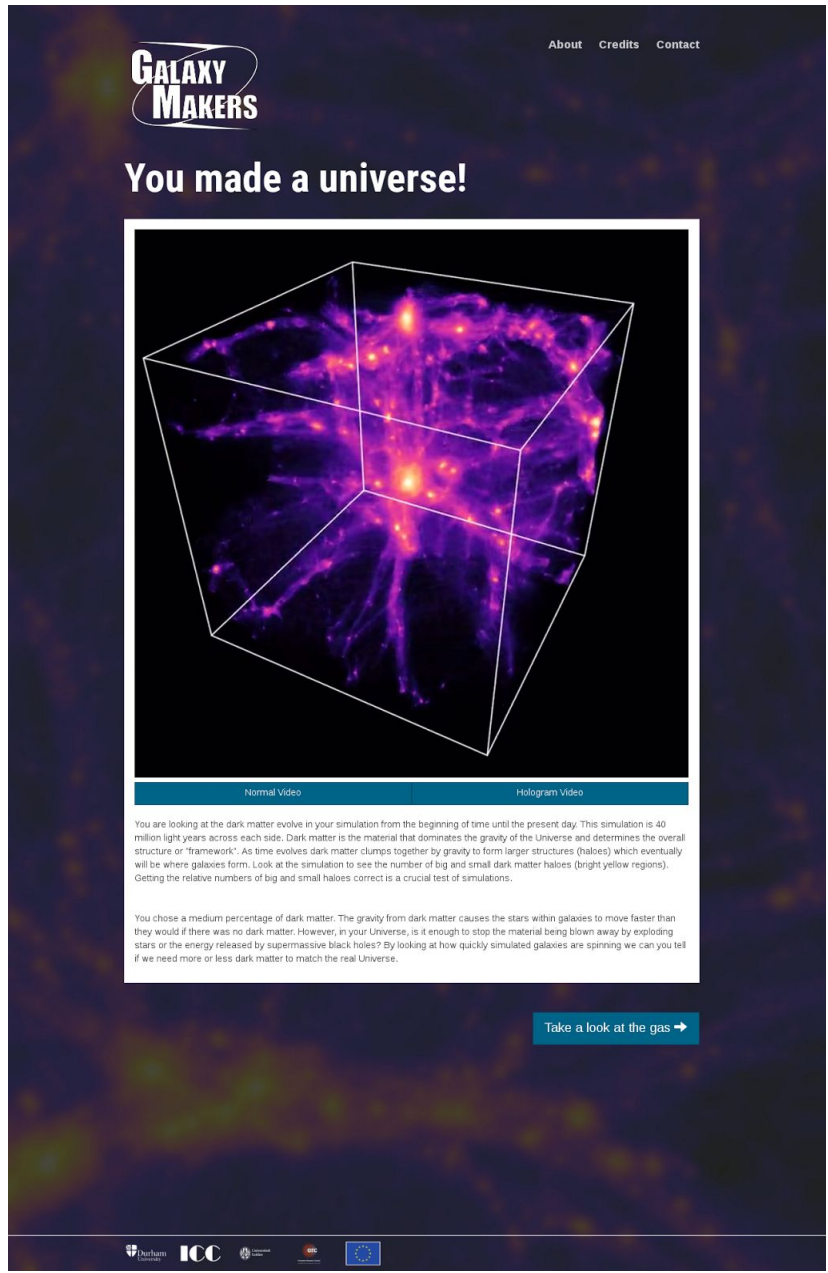

**Figure 4**: The Galaxy Makers website. This is the view that users are presented with after 'creating a universe'.

## The Power of Social Media

Social media is often over-appreciated in public engagement communities. It is quite clear that tweeting about research is no replacement for getting out there and directly engaging the public. This does not, however, prevent social media from being a highly useful tool.

Two months after the in-person exhibition had ended, an independent game developer found the Galaxy Makers website and tweeted it to his 5000 followers with the message "A wonderful lesson for all the #gamedev-s out there on how to gamify science." This tweet gave the website it's single largest traffic spike of all time (there were over 1100 engagements with the tweet) - including during the exhibition.

Even now, users of the website are still sharing their experiences on twitter and other forms of social media. Whilst some may be tired of the incessant cries of "use social media!" from public engagement professionals, interactions like these show us just how powerful social and new media are, and just how important it is to cultivate relationships on these websites. This sharing has allowed our content to be continually reused and allows it to remain visible, see Table 1 for metrics.

### Advertising

The website also proved useful during the in-person exhibits for advertising the exhibit to members of the public through social media. The ability to play around with an online experience before making the trip to the exhibition is certainly an interesting concept from an advertising point of view; this is a clear improvement over the traditional photos of people milling around the exhibition centre.

## Further Potential of Online Material

In some aspects, we did not fully utilise our opportunity to collect information about the exhibition through the online component. For example, we could have produced a questionnaire on the website for qualitative and quantitative feedback. Consequently we had a lack of participant evaluation, although some responses were collected through social media and email.

Another interesting possibility would have been to somehow include participant information in the code that was given out to attendees. This could have been used to track which groups were most interested in the website, and could have been interested by giving out slightly different codes to different groups (for example splitting by gender or age group) and tracking their use on the website. Including this information would have significantly bolstered our demographic data and is certainly a compelling opportunity for future exhibitions.

Social media could also have been used to greater effect during the Galaxy Makers exhibition. For example making the Twitter handle visible on all of our materials would have increased activity. Additionally, had the website received some attention from any high-profile social media users, then the engagement with the website would have certainly been higher. In the future, these relationships are something that we can work harder to cultivate.

## Conclusion

Much like some of the most interesting scientific breakthroughs, the success of the online exhibition came as a real surprise. The fact that nearly 40% of those that received a "hologram pyramid" went away and re-engaged with our content was astounding. Our expectation was that (maybe) 10% of users would go away and utilise the website. It is clear that the younger generation loves all things digital and that this is a productive route to go down to foster re-engagement.

The material created for Galaxy Makers is already having a much further-reaching impact than an in-person exhibition could have. We have been contacted by other universities who are using Galaxy Makers for their undergraduate labs, schoolteachers who are interested in using the project, and of course users stumbling across it on the internet.

When creating an exhibition, a parallel online experience is certainly something to consider. With only 1/25th of our total team working on the website, we have managed to re-engage hundreds of attendees and gain a large amount of valuable data for evaluation. We have also created a lasting, permanent home for our content that is now no longer relegated to a one-week experience; it is free to be shared around the world forever.

# Notes

1. More information about the EAGLE simulations and its creator are available here: http://icc.dur.ac.uk/Eagle/

# Acknowledgements

Many thanks to the builders of the EAGLE simulations (the Virgo Consortium) and the Galaxy Makers exhibit. In particular, we would like to extend our thanks to Professor Carlton Baugh for leading the Galaxy Makers project, and for his comments on this article. We would also like to thank thank the institutions that provided resources and funding either directly or indirectly: The Ogden Trust; The European Commission; Science and Technology Funding Council (including grant reference: ST/L00075X/1); Marie Curie Actions; The Royal Astronomical Society, the Royal Society and DiRAC.

| Metric | Exhibition (1st-17th of July) | Lifetime (to 1st December) |
|---|---|---|
| Sessions | 1580 | 3603 |
| Users | 892 | 2270 |
| Pageviews | 5639 | 12486 |
| Average Session Duration | 3:18 | 2:47 |
| Sessions by Return Users | 43.5% | 37.0% |

Table 1: Metrics have continued to grow over time, showing that the content produced for Galaxy Makers is continuing to be used. A large number of these interactions post-exhibition are due to users sharing their experiences on social media.